\documentclass[12pt,a4paper]{article}
\begin{document}

\begin{center}

\vspace{3cm}

{\bf \Large Infrared Safe Observables in\\[0.3cm]  ${\cal N}=4$ Super Yang-Mills
Theory}

\vspace{3cm}

{\bf L. V. Bork$^{\flat}$, D. I. Kazakov$^{\natural,\flat}$, G. S.
Vartanov$^{\natural,\sharp}$\footnote{Present address:
Max-Planck-Institut f\"ur Gravitationsphysik,
Albert-Einstein-Institut 14476 Golm, Germany}
and A. V. Zhiboedov$^{\natural,\|}$\footnote{Present address: Physics Department, Princeton University, Princeton NJ 08544, USA}}\vspace{0.5cm}\\
{\it $^\natural$ Bogoliubov Laboratory of Theoretical Physics, Joint
Institute for Nuclear Research, Dubna, Russia, \\
$^\flat$Institute for Theoretical and Experimental Physics, Moscow, Russia, \\
$^\sharp$University Center, Joint Institute for Nuclear Research, Dubna, Russia,\\
$^\|$Moscow State University, Physics Department, Moscow, Russia.
\\}

\abstract{The infrared structure of MHV gluon amplitudes in planar
limit for ${\cal N}=4$ super Yang-Mills theory is considered in the
next-to-leading order of PT. Explicit cancellation  of the infrared
divergencies in properly defined cross-sections is demonstrated. The
remaining finite parts for some inclusive differential
cross-sections in planar limit are calculated analytically. In
general, contrary to the virtual corrections, they do not reveal any
simple structure.}


PACS classification codes: 11.15.-q; 11.30.Pb; 11.25.Tq

\end{center}

\section{Introduction}

In recent years remarkable progress in understanding the structure of
planar\footnote{Defined as $g \rightarrow 0$; $N_c \rightarrow \infty$; $\lambda = g^2
N_c ~ fixed$ } ${\cal N}=4$ SYM (supersymmetric Yang-Mills) theory has been achieved. In
the planar limit this theory seems to  be the first example of solvable non-trivial
four-dimensional Quantum Field Theory.

The objects which were in the spotlight starting from AdS/CFT (Anti
de Sitter/Con\-for\-mal Field Theory) correspondence were local
operators, namely  the spectrum of their anomalous dimensions. They
were calculated on the one hand side from the field theory
approach~\cite{4-loop} and from the other side as the energy levels
of a string in classical background~\cite{GPK,FroTse03} revealing
remarkable coincidence.

Recently the quantities of interest are the so-called
MHV\footnote{MHV (maximally helicity violating) amplitudes are the
amplitudes where all particles are treated as outgoing and the net
helicity is equal to $n-4$ where $n$ is the number of particles. For
gluon amplitudes MHV amplitudes are defined as the amplitudes in
which all but two gluons have positive helicities} scattering
amplitudes. It happens that in the planar limit of ${\cal N}=4$ SYM
theory they have a truly simple structure~\cite{Parke,Mangano}. It
is useful to consider the color-ordered amplitude defined  through
the color decomposition
\begin{eqnarray}
\mathcal{A}_{n}^{(l-loop)}=g^{n-2}(\frac{g^2N_c}{16\pi^2})^{l}\sum_{perm}
Tr(T^{a(1)}...T^{a(n)})A^{(l)}_n(p_{a(1)},...,p_{a(n)}),
\end{eqnarray}
where $\mathcal{A}_{n}$ is the physical amplitude, $A_n$ are the  partial color-ordered
amplitudes, and $a_i$ is the color index of $i$-th external ``gluon''.

It was found that these amplitudes reveal  the iterative structure which was first
established in two loops~\cite{Anastasiou:2003kj} and then confirmed at the three loop
level by Bern, Dixon and Smirnov, who formulated the ansatz \cite{bds05} for the
$n$-point MHV amplitudes:
\begin{eqnarray}
&& \makebox[-2em]{} {\cal M}_n\equiv \frac{A_{n}}{A_{n}^{tree}} =
1+\sum\limits_{L=1}^\infty\left(\frac{\alpha}{4\pi}\right)^L\hspace{-0.2cm}
M_n^{(L)}(\epsilon)=\exp\left[\sum\limits_{l=1}^\infty\left(\frac{\alpha}{4\pi}\right)^l
\hspace{-0.15cm}
\left(f^{(l)}(\epsilon)
M_n^{(1)}(l\epsilon)\!+\!C^{(l)}\!+\!E^{(l)}_n(\epsilon)\right)\right],
\end{eqnarray}
where  $\alpha = g^2N_c/4\pi$ and $E_n^{(l)}$  vanishes as $\epsilon
\rightarrow 0$, $C^{(l)}$ are some finite constants, and
$M_n^{(1)}(l\epsilon)$ is the $l\epsilon$- regulated one-loop
n-point $\phi^3$ scalar amplitude.

It is not surprising that the IR divergent parts of the amplitudes factorize and
exponentiate \cite{infrared}. What is less obvious  is that it is also true for the
finite part
\begin{eqnarray}
&& \makebox[-2em]{} {\cal M}_n(\epsilon)=\exp\left[-\frac
18\sum\limits_{l=1}^\infty\left(\frac{\alpha}{4\pi}\right)^l
\left(\frac{\gamma^{(l)}_K}{(l\epsilon)^2}+\frac{2G^{(l)}_0}{l\epsilon}\right)
\sum\limits_{i=1}^n
\left(\frac{\mu^2}{-s_{i,i+1}}\right)^{l\epsilon}+\frac 14
\sum\limits_{l=1}^\infty\left(\frac{\alpha}{4\pi}\right)^l
\gamma^{(l)}_K F^{(1)}_n(0) \right],
\end{eqnarray}
where $\gamma_{K}$ is the so-called cusp anomalous dimension
\cite{colK} and $G_0$ is the second function (dependent on  the IR
regularization) which defines the IR structure of the amplitude
\cite{collinear} .

According to BDS ansatz the finite part of the amplitude is defined
by the cusp anomalous dimension and a function of kinematical
parameters specified at one-loop. For a four gluon amplitude one has
\begin{eqnarray}
 F^{(1)}_4(0)= \frac{\gamma_{K}}{4} \log^2 \frac{s}{t}.
\end{eqnarray}

The cusp anomalous dimension is a function of the gauge coupling,
for which the four terms of the weak coupling
expansion~\cite{4-loop} and two terms of the strong coupling
expansion~\cite{GPK,FroTse03} are known.  Integrability from the
both sides of the AdS/CFT correspondence leads  to all-order
integral equation~\cite{BES06} solution of which being expanded in
the coupling reproduces both the series~\cite{BaKo}.

While for $n=4,5$ the BDS ansatz goes through all checks, namely the
amplitudes were calculated up to four loops for four gluons
\cite{4-loop} and up to two loops for five gluons \cite{chazo}.
However, starting from $n=6$ it fails. The first indication of the
problem was strong coupling calculation in the limit $n \rightarrow
\infty$ \cite{am2} where discrepancy with the BDS formula was found.
The second indication came from the amplitude/Wilson loop duality
\cite{am1,Dks,Brandhuber,Dhks1} namely from the comparison of the
hexagonal light-like Wilson loop and finite part to the BDS ansatz
for the six-gluon amplitude. It was found that the two expressions
have difference in a non-trivial function of the three (dual)
conformally invariant variables \cite{Kor2l}. The third indication
appeared in the paper \cite{Lip} where the analytical structure of
the BDS ansatz was analyzed and starting from $n=6$ the Regge limit
factorization of the amplitude in some physical regions fails.
Finally it was shown by explicit two-loop calculation \cite{twoBern}
that the BDS ansatz is not true and it needs to be modified by some
unknown finite function, which  is an open and intriguing problem.
From two-loop calculation for the six-point amplitude \cite{twoBern}
and hexagonal light-like Wilson loop \cite{KorVer} it was shown that
the gluon amplitude/Wilson loop duality is still valid.

While all the UV divergences in ${\cal N}=4$ SYM are absent in scattering amplitudes the
IR ones remain and are supposed to be canceled in a properly defined quantities.
Regularized expressions act like some kind of scaffolding which has to be removed to
obtain eventual physical observable. It is these quantities that are the aim of our
calculation. And though the Kinoshita-Lee-Nauenberg~\cite{KLN} theorem in principle tell
us how to construct such quantities, explicit realization of this procedure is not simple
and one can think of various possibilities. In particular, one could consider the
so-called energy flow functions defined in terms of the energy-momentum tensor
correlators  introduced in \cite{EF} and considered  at weak coupling regime in
\cite{EFK} and recently  at strong coupling regime in \cite{hofmal}. From our side we
concentrated on inclusive cross-sections in the hope that they reveal some factorization
properties discovered in the regularized amplitudes. Similar questions were discussed in
\cite{vanNeerven:1985ja}, where the inclusive cross-section  like the IR safe observables
in $\cal{N}$=4 SYM were constructed.

\section{The Infrared Safe Observables}

To perform the procedure of cancellation of the IR divergences one
should have in mind that in conformal theory all the masses are zero
and one has additional collinear divergences which need special
care. In these work we employ the method developed in QCD parton
model \cite{EKSglu,EKS,KunsztSoper,Katani,Veretin}. It includes two
main ingredients in the cancellation of infrared divergencies coming
from the loops:  emission of additional soft real quanta and
redefinition of the asymptotic states resulting in the splitting
terms governed by the kernels of DGLAP equations. The latter ones
take care of the collinear divergences.

Typical observables in QCD parton model calculations are inclusive jet cross-sections,
where  the total energy of scattered partons is not fixed since they are considered to be
parts of the scattered hadrons. In~\cite{EKS} the algorithm of extracting divergences was
developed which allows one to cancel divergences and apply numerical methods for
calculation of a finite part. In our paper we choose as our observables the inclusive
cross-sections  with fixed initial energy and get analytical expression for the finite
part of the differential cross-section. We do not assume any confinement and consider the
scattering of the single parton based coherent states\footnote{squared perturbative
amplitudes used in our calculation have been summed over colors, so in this sense they
are colorless and there are no contradiction with statements that cancellation of IR
divergences occurs only for colorless objects.} being the asymptotic states of conformal
field theory.

When the number of particles increases one has to specify the measurable quantity and to
distinguish the particle(s) in the final state. If one wants to construct the finite
quantity it is not sufficient to consider the process with the fixed number of final
particles. One has to include processes with emission of additional soft and collinear
massless states, i.e to consider the inclusive cross-section.

One possibility  is to introduce the energy and angular resolution of the detector and to
cut the phase space so that the soft quanta with total energy below the threshold as well
as all the particles within the given solid angle are included. This procedure works well
in QED but introduces explicit dependence on the energy and angular cut off, thus
violating conformal invariance.

We adopt here the other attitude and do not introduce any cut off in the phase space but
rather consider the inclusive cross-section with emission of all possible particles
allowed by kinematics. Then one has to specify which particle is detected. For instance,
one can measure the scattering of a given particle on a given angle integrating over all
the other particles. As it will be clear later in this case one still cannot avoid
introducing some scale related to the definition of the asymptotic states of a theory.

\section{Calculation of Inclusive Cross-sections in ${\cal N}=4$ SYM theory}

Our aim is to evaluate the NLO correction to the inclusive differential polarized cross
section in the weak coupling limit in planar ${\cal N}=4$ SYM  in analytical form and to
trace the cancellation of the IR divergences.

We start with the $2 \to 2$ MHV scattering amplitude with two incoming positively
polarized gluons and two outgoing positively polarized gluons and consider the
differential cross-section $d\sigma_{2 \rightarrow 2}(g^{+}g^{+}\rightarrow
g^{+}g^{+})/d\Omega$ as a  function of the scattering solid angle. The total
cross-section is divergent at zero angle. Treating all the particles as outgoing this
amplitude is denoted as $(--,++)$ MHV amplitude. At tree level the cross-section is given
by
\begin{eqnarray}\label{tree}
  \frac{d \sigma_{2 \rightarrow 2}}{d\Omega_{13}} = \frac{1}{J} \int d\phi_{2} |{\cal M}_{4}^{(tree)}|^2\mathcal{S}_2,
\end{eqnarray}
where $J$ is flux factor, in our case $J=s$, $s$ is standard
Mandelstam variable, $d \phi_{2}$ is the phase volume of
two-particle process (we use dimensional regularization\footnote{
namely
 -- \textbf{FDH} version of dimentional reduction, see for details \cite{FDH}}, $D=4-2\epsilon$)
and $\mathcal{S}_n$ ($n=2$ in this case) is the so-called measurement function which
specifies what is really detected. $d\Omega_{13}=d\phi_{13}d
cos(\theta_{13})$\footnote{if to be more accurate in dimensional
regularization(reduction) we have
$d\Omega_{13}^{\epsilon}=d\phi_{13}sin(\phi_{13})^{-2\epsilon}d cos(\theta_{13})
sin(\theta_{13})^{-2\epsilon}$ .}, $\theta_{13}$ is the scattering angle of the particles
with momenta $\textbf{p}_3$ with respect to $\textbf{p}_1$  in the center of mass frame.
The matrix element is obtained from the color-ordered amplitudes via summation
\cite{Parke,Mangano}
\begin{equation}\label{4gluonel}
| {\cal M}_{4}^{(tree)(--++)}|^2 \ = \ g^4 N_c^2 (N_c^2-1)
\sum_{\sigma \in P_{3}} \frac{s_{12}^4}{s_{1 \sigma(1)}
s_{\sigma(1)\sigma(2)}s_{\sigma(2) \sigma(3)}s_{\sigma(3) 1}},
\end{equation}
where in all expressions  we take $s_{i j}=(p_{i}+p_{j})^2$.

Within dimensional regularization(reduction) the cross-section looks
like
\begin{eqnarray} &&
\hspace{-1.5cm}\left(\frac{d\sigma_{2 \rightarrow
2}}{d\Omega_{13}}\right)_0^{(--++)}\hspace{-0.3cm} =\frac{\alpha^2N_c^2}{2E^2}
 \left(\frac{s^4}{t^2u^2}+\frac{s^2}{t^2}+\frac{s^2}{u^2}\right)\left(\frac{\mu^2}{s}
 \right)^\epsilon
=\frac{\alpha^2N_c^2}{E^2}\left(\frac{\mu^2}{s}\right)^\epsilon\frac{4(3+c^2)}{(1-c^2)^2},
 \end{eqnarray}
where $s,t,u$ are the Mandelstam variables, $E$ is the total energy in the center of mass
frame, $c=\cos \theta_{13}$, $\mu$ and $\epsilon$ are the parameters of  the dimensional
regularization. The next step is to calculate the NLO corrections.

\subsection{Virtual part}

To get the one-loop contribution to the differential cross section
we use already known one loop  contribution for the color-ordered
amplitude~\cite{GSO}
$$M_4^{(1-loop)}(\epsilon)=A_4^{(1-loop)}/A_4^{(tree)}\ = - \frac 12 st I_4^{(1-loop)}(s,t),$$
where $I_4^{(1)}(s,t)$ is the scalar box diagram
$$ I_4^{(1-loop)}(s,t) \ = \
-\frac{2}{st}\frac{\Gamma(1\!+\!\epsilon)\Gamma(1\!-\!\epsilon)^2}{\Gamma(1-2\epsilon)}
[\frac{1}{\epsilon^2}\left((\frac{\mu^2}{s})^{\epsilon}\!\!+\!(\frac{\mu^2}{-t})^{\epsilon}\right)
\!+\!\frac{1}{2}\log^2\left(\frac{s}{-t}\right)\!+\!\frac{\pi^2}{2}]+\mathcal{O}(\epsilon).
$$

Then it is straightforward to obtain the one-loop contribution to
the cross-section in the planar limit
\begin{eqnarray}\label{1loop4gluon}
 && \hspace{-1.0cm}\left(\frac{d\sigma_{2 \rightarrow 2}}{d\Omega_{13}}\right)_{virt}^{(--++)}
 =\frac{\alpha^2 N_c^2}{E^2}
 \left(\frac{\mu^2}{s}\right)^{2\epsilon}4\left\{\frac{\alpha }{4\pi}
 \left[-\frac{16}{\epsilon^2}\frac{3+c^2}{(1-c^2)^2}
+ \frac{4}{\epsilon}
\left(\frac{5+2c+c^2}{(1-c^2)^2}\log(\frac{1-c}{2})\right.\right.\right.
\nonumber \\
 &&\hspace{-1cm}\left.\left.\left.+\frac{5-2c+c^2}{(1-c^2)^2}\log(\frac{1+c}{2})
\phantom{\frac{1+c}{2}}\hspace{-0.7cm}\right) +\frac{16(3+c^2)\pi^2}{3(1-c^2)^2}-
\frac{16}{(1-c^2)^2}\log(\frac{1-c}{2})\log(\frac{1+c}{2})\right]\right\}.
\end{eqnarray}
It should be  stressed that due to conformal  invariance of $\mathcal{N}=4$ SYM theory at
quantum level there are no UV divergences in (\ref{1loop4gluon}) and all divergences have
the IR or collinear nature. They have to cancel in the properly defined observables.

\subsection{Real emission}

The next step is the calculation of the amplitude with three outgoing particles. Here we
have to define which is the process that we are interested in. There are several
possibilities.
\begin{enumerate}
  \item Three gluons with positive helicities: $g^+g^+ \rightarrow g^+g^+g^+.$
  This is the MHV amplitude;
  \item Two gluons with positive helicities and the
  third one with negative helicity: $g^+g^+ \rightarrow
  g^+g^+g^-.$\footnote{There is also a $g^+g^+ \rightarrow
  g^+g^-g^+$ helicity configuration, but the partial amplitudes for them are equal. We will use
  $(--++-)$ notation for both of them.}
  This is the anti-MHV amplitude;
  \item One of three final particles is the gluon with positive
  helicity and the rest is the quark-antiquark pair\footnote{The $\mathcal{N}=4$
  supermultiplet
  consists of a gluon $g$, 4 fermions ("quarks") $q^{A}$ and 6 real scalars
  $\Lambda^{AB}$. $A$ and $B$ are $SU(4)_{R}$ indices, $\Lambda$ is an antisymmetric
  tensor. It is implied that all squared  amplitudes with quarks
  and scalars are summed over these indices.}:
  $g^+g^+ \rightarrow \ g^+q^-\overline{q}^+$ or $g^+g^+
  \rightarrow g^+q^+\overline{q}^-.$
  This  is an anti-MHV amplitude;
  \item One of three final particles is the gluon with positive
  helicity and the rest are  two scalars:
  $g^+g^+ \rightarrow g^+ \Lambda \Lambda.$ This  is an
  anti-MHV amplitude.
\end{enumerate}

If one fixes one gluon with positive helicity scattered at angle $\theta$ and sum over
all the other particles then all the processes mentioned above contribute.  In case when
one fixes two gluons with positive helicity and looks for the rest, only the first two
options are allowed.

The cross-section of these processes can be written as
\begin{eqnarray}\label{crossf}
 \frac{d \sigma_{2 \rightarrow 3}}{d\Omega_{13}} =  \frac{1}{J} \int d\phi_{3} | {\cal
 M}_5^{(tree)}|^2 \mathcal{S}_3,
\end{eqnarray}
where $d\phi_{3}$ is the three-particle phase volume and $\mathcal{S}_3$ is the
measurement function, which constraints the phase space and defines the particular
observable.

We omit  the details of the calculation and  for the lack of space present here only the
divergent parts of the calculated objects. We leave all the full answers for a separate
publication and present the finite part only in the simplest case below.

When there are three identical particles in the final state one has to define which ones
are detected. In case of one detectable particle one can choose the fastest one, in case
of two - the two fastest. Selection of detectable particles can be achieved restricting
the momentum. Thus choosing $p^0_3>E/3$ we fix the gluon with momentum $p_3$ as the
fastest particle.  When the final particles are not identical this problem does not
appear and the phase space is not restricted.  In what follows we  restrict the moment of
the gluon by a universal value $p^0_3> (1-\delta)/2 E$ and keep the value of $\delta$
arbitrary. The case of identical particles then corresponds to $\delta=1/3$ and the case
of nonidentical particles to $\delta=1$. One can show that in this case the requirements
of stability of observable with respect to emission of soft and collinear quanta
\cite{EKS} are satisfied. We show below that IR and collinear divergences cancel in
observables for an arbitrary value of $\delta$.

1. Real Emission (MHV)
\begin{eqnarray}
&&\makebox[-2em]{} \left(\frac{d\sigma_{2 \rightarrow
3}}{d\Omega_{13}}\right)^{(--+++)}_{Real} =\frac{\alpha^2N_c^2}{E^2}
 \left(\frac{\mu^2}{s}\right)^{2\epsilon}\frac{\alpha}{\pi}\left\{
\frac{1}{\epsilon^2}\frac{8(3\!+\!c^2)}{(1\!-\!c^2)^2}  +
\frac{1}{\epsilon}\left[
\frac{2}{(1\!+\!c)^2}\log(\frac{1\!-\!c}{2})+\frac{2}{(1\!-\!c)^2}\log(\frac{1\!+\!c}{2})
\right. \right. \nonumber \\
&& \left. \left. + \frac{16\delta(2\delta-3)}{(1-c^2)^2
(1-\delta)^2} +
\frac{12(3+c^2)}{(1-c^2)^2}\log\frac{1-\delta}{\delta}\right]+\mbox{Finite
part}\right\};
\end{eqnarray}
Notice the singularity when $\delta\to 1$.

2. Real Emission (anti-MHV)
\begin{eqnarray}
&&\hspace{-1.5cm}\left(\frac{d\sigma_{2 \rightarrow
3}}{d\Omega_{13}}\right)^{(--++-)}_{Real} =\frac{\alpha^2
N_c^2}{E^2}
 \left(\frac{\mu^2}{s}\right)^{2\epsilon}\frac{\alpha}{\pi}\left\{
\frac{1}{\epsilon^2}\frac{8(3+c^2)}{(1-c^2)^2}+\frac{1}{\epsilon}\left[
-\frac{12(c^2+3)\log\delta}{(1-c^2)^2} +
\frac{64 (12 c^2+17)}{3(1-c^2)^3}\right.\right. \nonumber\\
&&\hspace{-1.5cm}+\left. \left.  \frac{2\delta}{(1-c^2)^2}
\left( \frac 23 (5+3c^2) \delta^2 - (c^2+19) \delta + 2 (5c^2+43)
\right)+ \left(\frac{2(3c^2-24c+85)}{(1-c)(1+c)^3}\log(\frac{1-c}{2})
 \right. \right. \right.\nonumber\\
&&\hspace{-1.5cm}-\left.\left.\left.
\frac{8(c^2-6c+21)}{(1-c)(1+c)^3}\log(\frac{1+\delta\!-\!(1-\delta)c}{2})
-\frac{32(c^2-4c+7)}{(1+c)^3(1-c)(1+\delta-c(1-\delta))}
\right. \right. \right. \\
&& \hspace{-1.5cm}+ \left. \left.\left.
\frac{32(2-c)}{(1+c)^3(1+\delta-c(1-\delta))^2}
-\frac{64(1-c)}{3(1+c)^3(1+\delta-c(1-\delta))^3}+
(c\leftrightarrow -c) \right)\right]+ \mbox{Finite part}
\right\};\nonumber
\end{eqnarray}

3. Fermions
\begin{eqnarray}
&&\left(\frac{d\sigma_{2 \rightarrow
3}}{d\Omega_{13}}\right)^{(--+{\bar q} {q})}_{Real} =\frac{\alpha^2
N_c^2}{E^2}
 \left(\frac{\mu^2}{s}\right)^{2\epsilon}\frac{\alpha}{\pi}\left\{-
\frac{16}{\epsilon}\left[\frac{(79+25c^2)}{3(1-c^2)^2}\right.\right.
\\ \nonumber &&\hspace{-0.5cm}\left.\left.+
\frac{2(3-c)^2}{(1-c)(1+c)^3}\log(\frac{1-c}{2})
+\frac{2(3+c)^2}{(1-c)^3(1+c)}\log(\frac{1+c}{2})\right]+\mbox{Finite
part}\right\};
\end{eqnarray}

 4. Scalars
\begin{eqnarray}
&&\hspace{-0.5cm} \left(\frac{d\sigma_{2 \rightarrow
3}}{d\Omega_{13}}\right)^{(--+\Lambda\Lambda)}_{Real}
=\frac{\alpha^2 N_c^2}{E^2}
 \left(\frac{\mu^2}{s}\right)^{2\epsilon}\frac{\alpha}{\pi}\left\{-
\frac{8}{\epsilon}\left[-\frac{2(10+7c^2)}{(1-c^2)^2}\right.\right.
\\  \nonumber &&\hspace{-0.5cm}\left.\left.-
\frac{3(5-c)}{(1+c)^3}\log(\frac{1-c}{2})-\frac{3(5+c)}{(1-c)^3}
\log(\frac{1+c}{2})\right]
+\mbox{Finite part}\right\}.
\end{eqnarray}

\subsection{Splitting}

Taking into account emission of additional soft quanta allows one to cancel the IR
divergences (double poles in $\epsilon$) but leaves the single poles originating from
collinear ones. Indeed, in case of massless particles the asymptotic states (both the
initial and final ones) are not well defined since a massless quanta can split into two
parallel ones indistinguishable from the original. To take this into account one
introduces  the notion of distribution of the initial particle (gluon) with respect to
the fraction of the carried momentum $z$: $g(z)$. Then the initial distribution
corresponds to $g(z)=\delta(1-z)$, and the emission of a gluon leads to  a splitting: the
gluon carries the fraction of momentum equal $z$, while the collinear gluon - $(1-z)$.
The probability of this event is given by the so-called {\it splitting functions}
$P_{gg}(z)$. In case of a gluon in a final state this corresponds to the fragmentation of
the gluon into pair of gluons or pair of quarks or scalars.

Additional contributions from collinear particles in initial or final states to inclusive
cross-sections have the form, respectively
\begin{eqnarray}\label{AP}
d \sigma_{2 \rightarrow 2}^{spl,init} = \frac{\alpha}{2\pi}\frac{1}{\epsilon}
\left(\frac{\mu^2}{Q^2_f}\right)^\epsilon \!\!\int_{0}^{1}\!\! d z
P_{gg}(z)\!\!\!\!\!\sum\limits_{i,j=1,2;\ i\neq j}d \sigma_{2 \rightarrow 2} (z
p_{i},p_{j},p_{3},p_{4}){\cal S}_2^{spl,init}(z),
\end{eqnarray}
\begin{eqnarray}\label{APfin}
d \sigma_{2 \rightarrow 2}^{spl,fin} = \frac{\alpha}{2\pi}\frac{1}{\epsilon}
\left(\frac{\mu^2}{Q^2_f}\right)^\epsilon d \sigma_{2 \rightarrow 2}
(p_{1},p_{2},p_{3},p_{4}) \int_{0}^{1} d z\sum\limits_{l=g,q,\Lambda}P_{gl}(z) {\cal
S}_2^{spl,fin}(z),
\end{eqnarray}
where the scale $Q^2_f$, sometimes called the factorization scale,
belongs to the definition of the coherent asymptotic state and
restricts the value of transverse momenta.  The dependence of parton
distribution on $Q^2_f$ is governed by the DGLAP equation. The
splitting function $P_{ij}$ for each helicity configuration can be
obtained as a collinear limit of the corresponding partial amplitude
(see for example~\cite{Kunszt3Jet},~\cite{Dixon} fore more details).

 Taking into account the splitting of initial states and the fragmentation of
the final states we get the following contribution to the inclusive cross sections
\begin{enumerate}
  \item
The initial  and final splitting for the MHV amplitude.
\begin{eqnarray}
&&\left(\frac{d\sigma_{2 \rightarrow
3}}{d\Omega_{13}}\right)^{(--+++)}_{InSplit} =\frac{\alpha^2
N_c^2}{E^2}
 \left(\frac{\mu^2}{s}\right)^{\epsilon}\left(\frac{\mu^2}{Q_f^2}\right)^{\epsilon}\!\!
 \frac{\alpha}{\pi}\left\{
\frac{1}{\epsilon}\left[
-\frac{4(c^2\!+\!3)}{(1\!-\!c^2)^2}\left(\log\frac{1\!-\!c}{2}+\log\frac{1\!+\!c}{2}\right)
\right. \right. \nonumber \\ && \left. \left. \makebox[1em]{}
-\frac{8(c^2+3)}{(1-c^2)^2} \log\frac{1-\delta}{\delta}
-\frac{16\delta(2\delta-3)}{(1-c^2)^2(1-\delta)^2} \right]+
\mbox{Finite part}\right\},
\end{eqnarray}
\begin{eqnarray}
&& \hspace{-3cm}\left(\frac{d\sigma_{2 \rightarrow
3}}{d\Omega_{13}}\right)^{(--+++)}_{FnSplit} =\frac{\alpha^2
N_c^2}{E^2}
 \left(\frac{\mu^2}{s}\right)^{\epsilon}\left(\frac{\mu^2}{Q_f^2}\right)^{\epsilon}
 \frac{\alpha }{\pi}\left\{-\frac{1}{\epsilon}\frac{4(c^2+3)}{(1-c^2)^2}
\log\frac{1-\delta}{\delta} \right\};
\end{eqnarray}
  \item
The initial and final splitting for the anti-MHV amplitude
\begin{eqnarray}
&&\hspace{-1.8cm}\left(\frac{d\sigma_{2 \rightarrow
3}}{d\Omega_{13}}\right)^{(--++-)}_{InSplit} =\frac{\alpha^2
N_c^2}{E^2}
 \left(\frac{\mu^2}{s}\right)^{\epsilon}\hspace{-0.1cm}
 \left(\frac{\mu^2}{Q_f^2}\right)^{\epsilon}\hspace{-0.1cm}
 \frac{\alpha}{\pi}\left\{
\frac{1}{\epsilon}\left[\frac{8(c^2+3)}{(1-c^2)^2} \log\delta
- \frac{64 (12 c^2+17)}{3(1-c^2)^3}
\right.\right.  \nonumber \\
&&\hspace{-1.8cm}-\left.\left.
\frac{4\delta}{(1-c^2)^2} \left( \frac 23 (1+c^2) \delta^2 + (c^2-5) \delta +
2 (c^2+17) \right)+\left( \frac{4(c^3-15c^2+51c-45)}{(1-c)^2(1+c)^3}\log\frac{1\!-\!c}{2}
\right. \right.\right.\nonumber \\
&&\hspace{-1.8cm}+
\left.\left.  \left. \frac{8(c^2-6c+21)}{(1-c)(1+c)^3}
\log\frac{1+\delta-c(1-\delta)}{2}+
\frac{32(c^2-4c+7)}{(1+c)^3(1-c)(1+\delta-c(1-\delta))}
\right. \right. \right.\\
 && \hspace{-1.8cm}- \left. \left.\left.
\frac{32(2-c)}{(1+c)^3(1+\delta-c(1-\delta))^2}
+ \frac{64(1-c)}{3(1+c)^3(1+\delta-c(1-\delta))^3}+
(c\leftrightarrow -c) \right) \right]+ \mbox{Finite part}\right\},\nonumber
\end{eqnarray}
\begin{eqnarray}\hspace*{-1cm}
\left(\frac{d\sigma_{2 \rightarrow
3}}{d\Omega_{13}}\right)^{(--++-)}_{FnSplit}\hspace{-0.5cm}
&=&\frac{\alpha^2 N_c^2}{E^2}\!
 \left(\frac{\mu^2}{s}\right)^{\epsilon}\hspace{-0.2cm}
 \left(\frac{\mu^2}{Q_f^2}\right)^{\epsilon}\hspace{-0.2cm}
 \frac{\alpha }{\pi} \left\{\frac{1}{\epsilon}\frac{4(c^2+3)}{(1-c^2)^2}\left[
\log\delta\!-\!\delta(\frac 13\delta^2\!-\!\frac 32\delta\!+\!3)\right]\right\};
\end{eqnarray}
  \item
The initial splitting for the quark  final states ($\delta=1$)
\begin{eqnarray}
&&\hspace{-0.5cm} \left(\frac{d\sigma_{2 \rightarrow
3}}{d\Omega_{13}}\right)^{(--+\bar q q)}_{InSplit} =\frac{\alpha^2
N_c^2}{E^2}
 \left(\frac{\mu^2}{s}\right)^{\epsilon}\left(\frac{\mu^2}{Q_f^2}\right)^{\epsilon}
 \frac{\alpha }{\pi}\left\{
\frac{16}{\epsilon}\left[\frac{(79+25c^2)}{3(1-c^2)^2}\right.\right.
\\ \nonumber &&\hspace{-0.5cm}\left.\left.+
\frac{2(3-c)^2}{(1-c)(1+c)^3}\log(\frac{1-c}{2})
+\frac{2(3+c)^2}{(1-c)^3(1+c)}\log(\frac{1+c}{2})\right]+\mbox{Finite
part}\right\};
\end{eqnarray}
  \item
The initial splitting for the scalar final states ($\delta=1$)
\begin{eqnarray} &&\hspace{-0.5cm}
\left(\frac{d\sigma_{2 \rightarrow
3}}{d\Omega_{13}}\right)^{(--+\Lambda\Lambda)}_{InSplit}
=\frac{\alpha^2 N_c^2}{E^2}
 \left(\frac{\mu^2}{s}\right)^{\epsilon}\left(\frac{\mu^2}{Q_f^2}
 \right)^{\epsilon}\frac{\alpha }{\pi}\left\{
\frac{8}{\epsilon}\left[-\frac{2(10+7c^2)}{(1-c^2)^2}\right.\right. \\
&& \nonumber \hspace{-0.5cm}\left.\left.-
\frac{3(5-c)}{(1+c)^3}\log(\frac{1-c}{2})-\frac{3(5+c)}{(1-c)^3}
\log(\frac{1+c}{2})\right]
+\mbox{Finite part}\right\}.
\end{eqnarray}
\end{enumerate}

\section{IR safe observables in ${\cal N}=4$ SYM}

In the NLO there are  two sets of amplitudes, namely  the MHV and anti-MHV amplitudes,
which contribute to the observables.  The leading order  4-gluon amplitude is both MHV
and anti-MHV and we split it into two parts.  Then one can construct three types of
infrared-safe quantities in the NLO of perturbation theory, namely
\begin{itemize}
\item pure gluonic MHV amplitude
\begin{equation} \label{fin1} \hspace{-1cm}A^{MHV}=\frac
12\left(\frac{d\sigma_{2 \rightarrow
2}}{d\Omega_{13}}\right)^{(--++)}_{Virt}\hspace{-0.3cm}
+\left(\frac{d\sigma_{2 \rightarrow
3}}{d\Omega_{13}}\right)^{(--+++)}_{Real}\hspace{-0.3cm}+
\left(\frac{d\sigma_{2 \rightarrow
3}}{d\Omega_{13}}\right)^{(--+++)}_{InSplit}\hspace{-0.3cm}+
\left(\frac{d\sigma_{2 \rightarrow
3}}{d\Omega_{13}}\right)^{(--+++)}_{FnSplit};\end{equation}
\item pure gluonic anti-MHV amplitude
\begin{equation}  \label{fin2} \hspace{-1cm}B^{anti-MHV}=\frac
12\left(\frac{d\sigma_{2 \rightarrow
2}}{d\Omega_{13}}\right)^{(--++)}_{Virt}\hspace{-0.3cm}
+\left(\frac{d\sigma_{2 \rightarrow
3}}{d\Omega_{13}}\right)^{(--++-)}_{Real}\hspace{-0.3cm}+
\left(\frac{d\sigma_{2 \rightarrow
3}}{d\Omega_{13}}\right)^{(--++-)}_{InSplit}\hspace{-0.3cm}+
\left(\frac{d\sigma_{2 \rightarrow
3}}{d\Omega_{13}}\right)^{(--++-)}_{FnSplit};\end{equation}
\item anti-MHV amplitude with fermions or scalars fropm the full
$\mathcal{N} = 4$ supermultiplet
\begin{equation} \label{fin3}  \hspace{-1cm}C^{Matter}=
\left(\frac{d\sigma_{2 \rightarrow
3}}{d\Omega_{13}}\right)^{(--+,\ q{\bar q}+\Lambda\Lambda)}_{Real}+
\left(\frac{d\sigma_{2 \rightarrow
3}}{d\Omega_{13}}\right)^{(--+,\ q{\bar
q}+\Lambda\Lambda)}_{InSplit}
.\end{equation}
\end{itemize}
We would like to stress once more  that in each expression
(\ref{fin1},\ref{fin2},\ref{fin3}) \textit{all IR divergences cancel} for arbitrary
$\delta$ and only the finite part is left.

Defining now the physical condition for the observation we get several infrared-safe
inclusive cross-sections
\begin{itemize}
\item Registration of \textit{two fastest} gluons of positive helicity
\begin{equation} \label{Fin1} A^{MHV}\Big|_{\delta=1/3}+B^{anti-MHV}\Big|_{\delta=1};
\end{equation}
\item Registration of \textit{one fastest} gluon of positive helicity
\begin{equation} \label{Fin2} A^{MHV}\Big|_{\delta=1/3}+B^{anti-MHV}\Big|_{\delta=1/3}
+C^{Matter}\Big|_{\delta=1};
\end{equation}
\item
Anti-MHV cross-section
\begin{equation} \label{Fin3} B^{anti-MHV}\Big|_{\delta=1}+C^{Matter}\Big|_{\delta=1}.
\end{equation}
\end{itemize}

Relative simplicity of the virtual contribution (\ref{1loop4gluon}) which does not
contain any special functions but logs suggests similar structure of the real part.
However this is not the case. While the singular terms are simple enough and cancel
completely the finite parts are usually cumbersome and contain polylogarithms. The only
expression where they cancel corresponds to $\delta=1$ case which is possible only for
the last set of observables, namely for the anti-MHV cross-section (\ref{Fin3}). Choosing
the factorization scale to be $Q_f=E$ we get
\begin{eqnarray}
&&\hspace{-0.7cm}\left(\frac{d\sigma}{d\Omega_{13}}\right)_{AntiMHV}
=\frac{4\alpha^2 N_c^2}{E^2}\left\{\frac{3+c^2}{(1-c^2)^2}\right.  \\
&& \makebox[2em]{} \hspace{-1.2cm}\left.- \frac{\alpha }{4\pi}\left[ 2\frac{ (c^4\!+\!2
c^3\!+\!4 c^2\!+\!6 c\!+\!19) \log^2(\frac{1-c}{2})}{(1-c)^2 (1+c)^4}
 +2\frac{ (c^4\!-\!2 c^3\!+\!4 c^2\!-\!6 c\!+\!19)
\log^2(\frac{1+c}{2})}{(1-c)^4 (1+c)^2}\right.\right.  \nonumber \\
&& \makebox[2em]{} \hspace{-0.5cm}\left.\left. -8 \frac{(c^2+1)
\log(\frac{1+c}{2})\log(\frac{1-c}{2})}{(1-c^2)^2}
+\frac{6\pi^2(3c^2+13)-5 (61 c^2+99)}{9 (1-c^2)^2}\right.\right. \nonumber \\
&& \makebox[2em]{} \hspace{-0.5cm}\left.\left.-2\frac{(11 c^3\!-\!31 c^2\!-47 c\!-\!133)
\log(\frac{1-c}{2})}{3(1+c)^3 (1-c)^2} +2\frac{(11 c^3\!+\!31 c^2\!-\!47 c\!+\!133)
\log(\frac{1+c}{2})}{3(1-c)^3 (1+c)^2}\right]\right\}.\nonumber
\end{eqnarray}
One can see that even this expression does not repeat the  Born amplitude and does not
have any simple structure. Note, that the finite answer depends on the factorization
scale. This dependence comes from the asymptotic states which violate conformal
invariance of the Lagrangian. This dependence seems to be unavoidable  and reflects the
act of measurement. Construction of observables which do not contain any external scale
remains an open question.

\section{Discussion}

To solve the model might have different meaning. Calculation of divergences and
understanding of their structure is very useful but surely not enough. The knowledge of
the S-matrix would be the final goal though the definition of the S-matrix in conformal
theory is a problem. Even in the absence of the UV divergences there are severe IR
problems and matrix elements do not exist after removal of regularization. The experience
of QCD, which is very similar to $\mathcal{N}=4$ SYM theory from the point of view of the
IR problems, tells us that in inclusive cross-sections the IR divergences cancel and one
has finite physical observables. However, one either has to redefine the asymptotic
states or consider the scattering of  "hadrons". In both the cases one has to introduce
some parton distributions which are the functions of a fraction of momenta and, in higher
orders, of momenta transfer. This leads to appearance of a factorization scale which
breaks conformal invariance. This means that we do not have meaningful observables in a
pure conformal theory. The finite observables of the type considered here are the
inclusive cross-sections
\begin{eqnarray}
d\sigma^{incl}_{obs}&=&\sum\limits_{n=2}^{\infty}\int\limits_0^1\!dz_1q_1(z_1,
\frac{Q_f^2}{\mu^2})\! \int\limits_0^1\!dz_2q_2(z_2,
\frac{Q_f^2}{\mu^2})\prod\limits_{i=1}^{n}\int\limits_0^1\!dx_iq_i(x_i,
\frac{Q_f^2}{\mu^2})\times \\
   &&\hspace{-1cm}\times  d\sigma^{2\to n}(z_1p_1,z_2p_2,...)S_n(\{z\},\{x\})=
   g^4\sum\limits_{L=0}^\infty \ \left(\frac{g^2}{16\pi^2}\right)^L d\sigma_L^{Finite}
   (s,t,u,Q_f^2),\nonumber
\end{eqnarray}
which besides the kinematical variables contain the dependence on the factorization
scale.

Remarkable factorization properties of the MHV amplitudes accumulated in the BDS ansatz
(with the so far unknown modification) suggest the way of receiving "exact" results for
the amplitudes on shell. However, as we have already mentioned, it is the finite part
that we are really for. Unfortunately, our calculation has demonstrated that the simple
structure of the amplitudes governed by the cusp anomalous dimension has been totally
washed out by complexity of the real emission matrix elements integrated over the phase
space. This means that either $\mathcal{N}=4$ SYM theory  does not allow such a simple
factorizable solution or that we considered the unappropriate observables. Alternatively
one may relay on simplification in particular kinematic regime like the Regge limit where
the exponentiation is expected.

There is an interesting duality between the MHV amplitudes and the Wilson loop,  between
the weak and the strong coupling regime~\cite{am1,Dks,Brandhuber,Dhks1}. Probably it
would be possible using the AdS/CFT correspondence to construct the IR safe observables
at the strong coupling limit (similarly to what we did here) and to shed some light on
the "true" calculable objects in conformal theories.

\section*{Acknowledgements}

We would like to thank A.Gorsky, A.Kotikov, L. Lipatov, R. Roiban, A. Slavnov for
valuable discussions. In particular, we thank G.Korchemsky for useful comments and the
list of  references. Financial support from RFBR grant \# 08-02-00856 and grant of the
Ministry of Education and Science of the Russian Federation \# 1027.2008.2 is kindly
acknowledged. Two of us LB and GV are partially supported by the Dynasty foundation.


\begin{thebibliography}{99}

\bibitem{Parke} S.~J.~Parke and T.~R.~Taylor,
  \emph{An Amplitude for $n$ Gluon Scattering},
  Phys.\ Rev.\ Lett.\  {\bf 56}, 2459 (1986);\\
  F.~A.~Berends and W.~T.~Giele,
  \emph{Recursive Calculations for Processes with n Gluons},
  Nucl.\ Phys.\  B {\bf 306}, 759 (1988).

\bibitem{GPK} S.S.~Gubser, I.R.~Klebanov and A.M.~Polyakov,
\emph{A semi-classical limit of the gauge/string correspondence},
Nucl.\ Phys.\  B {\bf 636} (2002) 99 [arXiv:hep-th/0204051].

\bibitem{FroTse03}
S.~Frolov and A.A.~Tseytlin, \emph{Semiclassical quantization of
rotating superstring in AdS(5) $X$ S(5)}, JHEP 
{\bf 0206} (2002) 007 [arXiv:hep-th/0204226].

\bibitem{Mangano}  M.~L.~Mangano and S.~J.~Parke,\emph{
Multiparton amplitudes in gauge theories}, Phys.\ Rept.\  {\bf 200},
301 (1991)  [arXiv:hep-th/0509223].

\bibitem{Anastasiou:2003kj}
  C.~Anastasiou, Z.~Bern, L.~J.~Dixon and D.~A.~Kosower,
  \emph{Planar amplitudes in maximally supersymmetric Yang-Mills theory},
  Phys.\ Rev.\ Lett.\  {\bf 91} (2003) 251602
  [arXiv:hep-th/0309040].

\bibitem{bds05} Z.~Bern, L.~J.~Dixon and V.~A.~Smirnov,
  \emph{Iteration of planar amplitudes in maximally supersymmetric Yang-Mills
  theory at three loops and beyond},
  Phys.\ Rev.\  D {\bf 72} (2005) 085001
  [arXiv:hep-th/0505205].

\bibitem{infrared} A.~H.~Mueller,
  \emph{On The Asymptotic Behavior Of The Sudakov Form-Factor},
  Phys.\ Rev.\  D {\bf 20} (1979) 2037;
\\
  J.~C.~Collins,
  \emph{Algorithm To Compute Corrections To The Sudakov Form-Factor},
  Phys.\ Rev.\  D {\bf 22} (1980) 1478;
\\
  A.~Sen,
  \emph{Asymptotic Behavior Of The Sudakov Form-Factor In QCD},
  Phys.\ Rev.\  D {\bf 24} (1981) 3281;
\\
  G.~P.~Korchemsky,
  \emph{Double logarithmic asymptotics in QCD},
  Phys.\ Lett.\  B {\bf 217} (1989) 330;
\\
  L.~Magnea and G.~Sterman,
  \emph{Analytic continuation of the Sudakov form-factor in QCD},
  Phys.\ Rev.\  D {\bf 42} (1990) 4222;
\\
  G.~P.~Korchemsky,
  \emph{Sudakov Form-Factor In QCD},
  Phys.\ Lett.\  B {\bf 220} (1989) 629.

\bibitem{colK} S.~V.~Ivanov, G.~P.~Korchemsky and A.~V.~Radyushkin,
\emph{Infrared Asymptotics Of Perturbative QCD: Contour Gauges},
Yad.\ Fiz.\  {\bf 44} (1986) 230 [Sov.\ J.\ Nucl.\ Phys.\  {\bf 44}
(1986) 145]. \\ G.~P.~Korchemsky and A.~V.~Radyushkin, \emph{Loop
Space Formalism And Renormalization Group For The Infrared
Asymptotics Of QCD}, Phys.\ Lett.\  B {\bf 171} (1986) 459.

\bibitem{collinear} L.~J.~Dixon, L.~Magnea and G.~Sterman,
  \emph{Universal structure of subleading infrared poles in gauge theory
  amplitudes},
  [arXiv:0805.3515 [hep-ph]];\\
  L.~F.~Alday,
  \emph{Universal structure of subleading infrared poles at strong coupling},
  [arXiv:0904.3983 [hep-th]].

\bibitem{4-loop} Z.~Bern, M.~Czakon, L.J.~Dixon, D.A.~Kosower and V.A.~Smirnov,
\emph{The Four-Loop Planar Amplitude and Cusp Anomalous Dimension in
Maximally Supersymmetric Yang-Mills Theory}, Phys.\ Rev.\  D {\bf
75} (2007) 085010  [arXiv:hep-th/0610248];\\
F.~Cachazo, M.~Spradlin and A.~Volovich, \emph{Four-Loop Cusp
Anomalous Dimension From Obstructions}, Phys.\ Rev.\  D {\bf 75}
(2007) 105011 [arXiv:hep-th/0612309].

\bibitem{BES06}
B.~Eden and M.~Staudacher, \emph{Integrability and
transcendentality}, J.\ Stat.\ Mech.\  {\bf 0611} (2006) P014
[arXiv:hep-th/0603157].

N.~Beisert, B.~Eden and M.~Staudacher, \emph{Transcendentality and
crossing}, J.\ Stat.\ Mech.\  {\bf 0701} (2007)
P021[arXiv:hep-th/0610251].

\bibitem{BaKo}
B.~Basso, G.~P.~Korchemsky, J.~Kotanski, \emph{Cusp anomalous
dimension in maximally supersymmetric Yang-Mills theory at strong
coupling}, Phys.\ Rev.\ Lett.\ {\bf 100}:091601, (2008).

\bibitem{chazo}
  F. Cachazo, M. Spradlin and A. Volovich, \emph{Iterative structure within the five-particle two-loop amplitude}, Phys. Rev. D 74, 045020 (2006)
 [arXiv:hep-th/0602228];\\
 Z. Bern, M. Czakon, D. A. Kosower, R.Roiban and V. A. Smirnov, \emph{Two-Loop Iteration of Five-Point $\mathcal{N}$=4 Super-Yang-Mills Amplitudes}, Phys. Rev. Lett. 97, 181601 (2006)
 [arXiv:hep-th/0604074].

\bibitem{am2}
  L.~F.~Alday and J.~Maldacena,\emph{
  Comments on gluon scattering amplitudes via AdS/CFT},
  JHEP {\bf 0711} (2007) 068
  [arXiv:0710.1060 [hep-th]].

\bibitem{am1}
  L.~F.~Alday and J.~M.~Maldacena,
  \emph{Gluon scattering amplitudes at strong coupling},
 JHEP {\bf 0706} (2007) 064
  [arXiv:0705.0303 [hep-th]].

\bibitem{Dks}
  J.~M.~Drummond, G.~P.~Korchemsky and E.~Sokatchev,
  \emph{Conformal properties of four-gluon planar amplitudes and Wilson loops},
Nucl.\ Phys.\  B {\bf 795} (2008) 385
  [arXiv:0707.0243 [hep-th]].

\bibitem{Brandhuber}
  A.~Brandhuber, P.~Heslop and G.~Travaglini,
  \emph{MHV Amplitudes in N=4 Super Yang-Mills and Wilson Loops},
  Nucl.\ Phys.\  B {\bf 794} (2008) 231
  [arXiv:0707.1153 [hep-th]].

\bibitem{Dhks1}
  J.~M.~Drummond, J.~Henn, G.~P.~Korchemsky and E.~Sokatchev,
  \emph{On planar gluon amplitudes/Wilson loops duality},
  Nucl.\ Phys.\  B {\bf 795} (2008) 52
  [arXiv:0709.2368].

\bibitem{Kor2l}
  J.~M.~Drummond, J.~Henn, G.~P.~Korchemsky and E.~Sokatchev,
  \emph{The hexagon Wilson loop and the BDS ansatz for the six-gluon amplitude},
  Phys.\ Lett.\  B {\bf 662} (2008) 456
  [arXiv:0712.4138 [hep-th]].

\bibitem{Lip} J.~Bartels, L.~N.~Lipatov, A.~S.~Vera,\emph{
  BFKL Pomeron, Reggeized gluons and Bern-Dixon-Smirnov amplitudes},
  [arXiv:0802.2065 [hep-th]].

\bibitem{twoBern}  Z.~Bern, L.~J.~Dixon, D.~A.~Kosower, R.~Roiban, M.~Spradlin, C.~Vergu and A.~Volovich,
  \emph{The Two-Loop Six-Gluon MHV Amplitude in Maximally Supersymmetric Yang-Mills
  Theory},
  [arXiv:0803.1465 [hep-th]].

\bibitem{KorVer}
  J.~M.~Drummond, J.~Henn, G.~P.~Korchemsky and E.~Sokatchev,
  \emph{Hexagon Wilson loop = six-gluon MHV amplitude},
  Nucl.\ Phys.\  B {\bf 815} (2009) 142
  [arXiv:0803.1466 [hep-th]].

\bibitem{KLN} T.~Kinoshita,
  \emph{Mass singularities of Feynman amplitudes},
  J.~Math.~Phys. {\bf 3} (1962) 650;\\
  T.~D.~Lee, M.~Nauenberg,
  \emph{Degenerate Systems and Mass Singularities},
  Phys.~Rev. {\bf 133} (1964) B1549.

\bibitem{EF}
  N.~A.~Sveshnikov and F.~V.~Tkachov,
  \emph{ Jets and quantum field theory},
  Phys.\ Lett.\  B {\bf 382} (1996) 403
  [arXiv:hep-ph/9512370]. \\   M.~Testa,
  \emph{Exploring the light-cone through semi-inclusive hadronic
  distributions},
  JHEP {\bf 9809} (1998) 006
  [arXiv:hep-ph/9807204].

\bibitem{EFK}
  G.~P.~Korchemsky, G.~Oderda and G.~Sterman,
  \emph{Power corrections and nonlocal operators},
  arXiv:hep-ph/9708346.\\
  G.~P.~Korchemsky and G.~Sterman,
  \emph{Power corrections to event shapes and factorization},
  Nucl.\ Phys.\  B {\bf 555}, 335 (1999)

\bibitem{hofmal} D.~M.~Hofman and J.~Maldacena,
 \emph{Conformal collider physics: Energy and charge correlations},
  JHEP {\bf 0805}, 012 (2008)
  [arXiv:0803.1467 [hep-th]].

\bibitem{vanNeerven:1985ja}
  W.~L.~van Neerven,
  \emph{Infrared Behavior Of On-Shell Form-Factors In A N=4 Supersymmetric
  Yang-Mills Field Theory},
  Z.\ Phys.\  C {\bf 30}, 595 (1986).

\bibitem{EKSglu} S.~D.~Ellis, Z.~Kunszt, D.~E.~Soper,
  \emph{The One Jet Inclusive Cross-Section at Order $\alpha_s^3$. 1. Gluons
  Only},
  Phys.\ Rev.\ D {\bf 40}:2188,1989.

\bibitem{EKS} S.~D.~Ellis, Z.~Kunszt, D.~E.~Soper,
\emph{The One Jet Inclusive Cross-Section at Order $\alpha_s^3$.
Quarks and gluons},
  Phys.~Rev.~Lett. {\bf 64} (1990) 2121.

\bibitem{KunsztSoper} Z.~Kunszt, D.~E.~Soper,
 \emph{ Calculation of jet cross-sections in hadron collisions at order $\alpha_s^3$},
  Phys.~Rev.~D {\bf 46}~(1992) 192.

\bibitem{Katani} S.~Catani, M.~H.~Seymour,
\emph{A General algorithm for calculating jet cross-sections in NLO
QCD}, Nucl.\ Phys.\ B{\bf 485} (1997) 291 [hep-ph/9605323].

\bibitem{Veretin} B.~A.~ Kniehl, A.~V.~ Kotikov, Z.~V.~ Merebashvili, O.~L.~
   Veretin,
    \emph{Heavy-quark pair production in polarized photon-photon collisions at next-to-leading order:
   Fully integrated total cross sections},
   Phys.\ Rev.\ D {\bf 79}:114032,2009 [arXiv:0905.1649 [hep-ph] ].

\bibitem{FDH} Z.~Bern, D.~A.~Kosower, \emph{The Computation of loop amplitudes in gauge
theories}, Nucl.\ Phys.\ B {\bf 379} (1992) 451.

\bibitem{GSO} M.~B.~Green, J.~H.~Schwarz and L.~Brink,
\emph{N=4 Yang-Mills And N=8 Supergravity As Limits Of String Theories}
  Nucl.\ Phys.\ B {\bf 198}, 474 (1982).

\bibitem{Kunszt3Jet}S. Frixione, Z. Kunszt, A. Signer,
\emph{Three-jet cross sections to next-to-leading order}, Nucl.\
Phys.\  B {\bf 467} (1996) 399. [hep-ph/9703305].

\bibitem{Dixon} L.~J.~Dixon,
\emph{Calculating scattering amplitudes efficiently},
[arXiv:hep-ph/9601359].

\end{thebibliography}
\end{document}